\documentclass[10pt,doublecolumn,twoside]{IEEEtran}

\usepackage{etoolbox}
\newtoggle{doublecolumn}
\newtoggle{calculateFigures}

\toggletrue{doublecolumn}
\togglefalse{calculateFigures}
\usepackage{etex}

\usepackage[english]{babel}
\usepackage{lipsum}
\usepackage{amsbsy,amsmath,amsfonts,amssymb,amsthm}
\usepackage{mathtools}
\usepackage{textcomp} 
\usepackage{relsize}

\usepackage{bm,cite,cases,pstricks,times,url,verbatim} 
\usepackage[noend]{algpseudocode}
\usepackage{booktabs}
\usepackage{tabularx,array,dcolumn,multirow}

\usepackage{dutchcal} 

\usepackage{soul} 

\usepackage{blkarray,bigdelim} 

\allowdisplaybreaks[4]

\usepackage{centernot} 

\newcommand{\subparagraph}{}

\iftoggle{doublecolumn}{
    \newtheorem{thm}{Theorem}
    \newtheorem{fact}{Fact}
    \newtheorem{lemma}{Lemma}
    \newtheorem{definition}{Definition}
    \newtheorem{conj}{Conjecture}
    \newtheorem{propos}{Proposition}
    \newtheorem{corol}{Corollary}
    \newtheorem{ass}{Assumption}
    \newtheorem{example}{Example}
    \newtheorem{remark}{Remark}
    \newtheorem{note}{Note}
    \newtheorem{obs}{Observation}
}{

    \newtheoremstyle{exampstyle}
      {0} 
      {0} 
      {\itshape} 
      {} 
      {\bfseries} 
      {.} 
      {.5em} 
      {} 

    \theoremstyle{exampstyle} 
    \theoremstyle{exampstyle} 
    \theoremstyle{exampstyle} \newtheorem{lemma}{Lemma}
    \theoremstyle{exampstyle} 
    \theoremstyle{exampstyle} 
    \theoremstyle{exampstyle} \newtheorem{propos}{Proposition}
    \theoremstyle{exampstyle} 
    \theoremstyle{exampstyle} 
    \theoremstyle{exampstyle} 
    \theoremstyle{exampstyle} 
    \theoremstyle{exampstyle} 
    \theoremstyle{exampstyle} 
}

\makeatletter
\newcommand{\pushright}[1]{\ifmeasuring@#1\else\omit\hfill$\displaystyle#1$\fi\ignorespaces}
\newcommand{\pushleft}[1]{\ifmeasuring@#1\else\omit$\displaystyle#1$\hfill\fi\ignorespaces}

\begingroup
\catcode`\#=11
\gdef\noautorotate{-dAutoRotatePages#/None}
\endgroup

\usepackage{graphicx,xcolor,float,dblfloatfix}
\usepackage{psfrag}

\usepackage{caption}
\usepackage{subcaption}

\newcommand{\subalign}[1]{%
  \vcenter{%
    \Let@ \restore@math@cr \default@tag
    \baselineskip\fontdimen10 \scriptfont\tw@
    \advance\baselineskip\fontdimen12 \scriptfont\tw@
    \lineskip\thr@@\fontdimen8 \scriptfont\thr@@
    \lineskiplimit\lineskip
    \ialign{\hfil$\m@th\textstyle##$&$\m@th\textstyle{}##$\crcr
      #1\crcr
    }%
  }
}

\iftoggle{calculateFigures}{
    \usepackage[crop=off,runs=2,pspdf=\noautorotate]{auto-pst-pdf}
}{
    \usepackage[off]{auto-pst-pdf}
}

\usepackage{hyperref}

\graphicspath{%
{./Figures/}
}

\usepackage{caption}
\usepackage{subcaption}
\begin{document}

\author{\IEEEauthorblockN{Alessandro~Biason,~\IEEEmembership{Student~Member,~IEEE,} and~Michele~Zorzi,~\IEEEmembership{Fellow,~IEEE}
\thanks{The authors are with the Dept. of Information Engineering, University of
Padova, Padova, Italy. email: \{biasonal,nil,zorzi\}@dei.unipd.it.}
\thanks{A preliminary version of this paper will be presented at IEEE WCNC 2016~\cite{Biason2016a}.}
}}

\title{Battery-Powered Devices in WPCNs}

\maketitle

\begin{abstract}
Wireless powered communication networks are becoming an effective solution for improving self sustainability of mobile devices. In this context, a hybrid access point transfers energy to a group of nodes, which use the harvested energy to perform computation or transmission tasks. While the availability of the wireless energy transfer mechanism opens up new frontiers, an appropriate choice of the network parameters (\emph{e.g.}, transmission powers, transmission duration, amount of transferred energy, etc.) is required in order to achieve high performance. In this work, we study the throughput optimization problem in a system composed of an access point which recharges the batteries of two devices at different distances. 
In the literature, the main focus so far has been on \emph{slot-oriented} optimization, in which all the harvested energy is used in the same slot in which it is harvested. However, this approach is strongly sub-optimal because it does not exploit the possibility to store the energy and use it at a later time. Thus, instead of considering the slot-oriented case, we address the \emph{long-term} maximization. This assumption greatly increases the optimization complexity, requiring to consider, \emph{e.g.}, the channel state realizations, its statistics and the batteries evolution. Our objective is to find the best scheduling scheme, both for the energy transferred by the access point and for the data sent by the two nodes. We discuss how to perform the maximization with optimal as well as approximate techniques and show that the slot-oriented policies proposed so far are strongly sub-optimal in the long run.
\end{abstract}


\section{Introduction}\label{sec:introduction}

\IEEEPARstart{N}{ew} generation devices, \emph{e.g.}, Wireless Sensor Networks (WSNs) or mobile cellular networks, are able to provide high communication performance in terms of throughput or delay at the cost of computational complexity and demanding power supplies. Wireless Energy Transfer (WET) has been recognized as one of the most appealing solutions for supplying mobile devices when their batteries cannot be easily or cheaply replaced. Via WET it becomes possible to greatly extend the network lifetime and improve the devices performance by avoiding energy outage situations. Nowadays, it is possible to transfer powers of tens to hundreds of microwatts at distances of $10$~m and $5$~m (see, for example, the Powercast company products~\cite{Powercast}) and thus it becomes  possible to supply ultra-low power mid-range networks. Differently from standard ambient energy harvesting techniques, WET has the major advantage of being fully controlled and does not rely on an external random phenomenon.

We consider a Wireless Powered Communication Network (WPCN) in which a base station transfers energy to different users. In WPCNs, one of the main goals is to design the scheduling procedure in order to better exploit the available resources and meet some Quality of Service (QoS) criteria (\emph{e.g.}, delay, packet drop rate, throughput). When a node is far away from the base station, it experiences a worse channel than the closer devices, on average, both for data transmission in uplink and for energy harvesting in downlink. Therefore, in order to develop a fair system in which all users achieve the same QoS, more resources have to be used to feed the far users. This phenomenon is known in the literature as \emph{doubly near-far effect} and has been solved in the slot-oriented case (in which all the transferred energy is also used for data transmission in the same time slot)~\cite{Ju2014}. However, in the battery-powered case, in which the harvested energy can be stored and used at a later time, new considerations can be made (channel condition, current battery levels, battery sizes, future energy arrivals, etc.) and the optimization approach becomes more involved. The goal of the present work is to investigate such a problem.

Wireless Energy Transfer techniques have experienced a renewed research interest in the last few years~\cite{Xiao2015} and several applications can be found in the WSN field, where low-power devices are fed with the transferred energy and use it for transmission or computation purposes. Different aspects of WET have been studied by both industry and academia, \emph{e.g.}, in terms of circuit and rectenna design~\cite{Nintanavongsa2012} but also in terms of transmission protocols by the communication and networking community. In this field, three major research areas can be identified so far: SWIPT, energy cooperation and WPCN. 
In SWIPT (Simultaneous Wireless Information and Power Transfer) systems, the tradeoffs between information and energy transfer are investigated~\cite{Grover2010}. Nowadays, because of hardware constraints of the current technology, a real ``simultaneous'' data and energy transmission is not possible yet, and therefore the Time Splitting (TS) and Power Splitting (PS) approaches were introduced~\cite{Zhang2013}. The TS approach, in which WET and data transmission are temporally interleaved, was studied in~\cite{Liu2013,Krikidis2012}, whereas PS was analyzed in~\cite{Timotheou2014,Nasir2013,Park2013a,Shi2014}. 
A second research area studies the energy cooperation paradigm, where different nodes exchange their energy to improve the system performance. This is particularly suitable for achieving energy fairness among devices when one has more energy resources (\emph{e.g.}, it is recharged by an external and powerful ambient energy source). The concept of energy cooperation was introduced in~\cite{Gurakan2013}, in which Gurakan~\emph{et al.} studied a system of a few nodes and defined the optimal offline communication schemes. Recently, \cite{Ni2015}~studied a similar system in which nodes receive energy by the same external energy source and introduced a save-then-transmit scheme. \cite{Tutuncuoglu2015a}~analyzes a multiterminal network with energy harvesting nodes which transfer and receive energy from other devices in the network. \cite{Biason2015e}~introduced achievable performance upper bounds for a transmitter-receiver case with finite energy buffers with or without energy cooperation. In~\cite{Gurakan2015}, routing with energy cooperation was studied.
Even if the previous research topics have mainly been studied separately, it is expected that in the near future a system which considers multiple aspects of WET will be analyzed or developed.

While several different kinds of WET mechanisms are available, \emph{e.g.}, inductive coupling or strongly coupled magnetic resonances~\cite{Kurs2007,Sample2011}, in this work we focus on Radio-Frequency  WET (RF-WET). Indeed, since RF-WET is a far-field WET technique, it is suitable for powering several devices simultaneously in a distributed area.
Via dedicated components, namely \emph{rectifiers}~\cite{Dolgov2010} (which, for example, can be composed of a diode~\cite{Hagerty2004}, a bridge of diodes or a voltage rectifier multiplier), the devices are able to convert the input RF signal into DC voltage, which can be used to refill their batteries. The RF signal can be harvested from the environment (\emph{e.g.}, this may be possible in a city where several electromagnetic sources are available), or from a dedicated source, \emph{i.e.}, a particular node (generally the access point) which emits RF signals to feed the devices (commercial products for RF-WET transmission/reception are already available, see~\cite{Powercast}). This last kind of scenario is known as \emph{wireless powered communication network} (WPCN).

In a WPCN where multiple devices harvest energy from the base station and transmit data in uplink, a doubly near-effect phenomenon is present: a user far away from the base station experiences, on average, a worse channel than the others both in uplink (therefore it has to use more energy to perform its transmission) and in downlink (thus it gathers less energy). The \emph{doubly near-far} problem was initially studied in~\cite{Ju2014}. The authors introduced a ``harvest-then-transmit'' scheme in which the time horizon is divided in slots and every slot is divided in two phases: first, the access point transfers energy to the devices and, secondly, the devices use the harvested energy to transmit data in the uplink channel. The trade-offs between the times to use for transferring energy and transmitting data were investigated and the optimal scheduling scheme was provided. The authors extended their previous work in~\cite{Ju2014a}, where user cooperation was taken into account in a two-device system. It was shown that coordination is a powerful technique which can effectively improve the system performance. Nevertheless, because of the additional complexity demanded to compute the scheduling scheme, and the unavoidable coordination and physical proximity required among devices, the cooperation solution may not be suitable for every scenario.
\cite{Chen2015}~described a ``harvest-then-cooperate'' protocol, in which source and relay work cooperatively in the uplink phase for the source's information transmission. The authors also derived an approximate closed-form expression for the average throughput of the proposed protocol.
\cite{Kim2015}~studied the case of devices with energy and data queues and described a Lyapunov approach to derive the stochastic optimal control algorithm which minimizes the expected energy downlink power and stabilizes the queues. The long-term performance of a single-user system for a simple transmission scheme was presented in closed form in~\cite{Morsi2014}. \cite{Hoang2014}~modeled a WPCN with a Decentralized Partially Observable Markov Decision Process (Dec-POMDP) and minimized the total number of waiting packets in the network. 
Similarly to~\cite{Ju2014}, a WPCN was studied in~\cite{Liu2014}, where the access point has also the capability of beamforming the transferred RF signal in order to serve the most disadvantaged users and to guarantee throughput fairness. The authors managed to convert a non-convex optimization problem into a spectral radius minimization problem, which can efficiently solved. \cite{Yang2015}~studied the applicability of the massive multiple-input-multiple-output (MIMO) technology to a WPCN. With massive MIMO it becomes possible to receive data from several different devices simultaneously (thanks to spatial multiplexity) but also to improve the downlink performance by using sharp beams. Most previous works describe a half-duplex system in which uplink and downlink cannot be performed simultaneously. Instead, the full duplex case was studied in~\cite{Ju2014b,Kang2014}. \cite{Ju2014b}~optimized the time allocations for WET and data transmission for different users in order to maximize the weighted sum rate of the uplink transmission. The authors considered perfect as well as imperfect self-interference (SIC) at the access point and showed that, when SIC is performed effectively, the performance of the full-duplex case outperforms the half-duplex one. A survey of recent advances and future perspectives in the WPCN field can be found in~\cite{Bi2015}.

In this work, we consider a WPCN composed of an Access Point (AP) and two distributed nodes. AP transfers energy in downlink to the nodes, which use the harvested energy for transmission purposes. Our system model is similar to that of~\cite{Ju2014,Liu2014,Morsi2014}. As in~\cite{Morsi2014}, we consider \emph{battery-powered} devices and focus on the long-term performance. However, \cite{Morsi2014}~considered only one device, whereas, in the present work, we consider the near-far effect problem when multiple devices are present. Moreover, differently from~\cite{Morsi2014}, we describe how to derive the optimal strategy to maximize the throughput of the system, whereas~\cite{Morsi2014} focuses on the performance evaluation of a given strategy. \cite{Ju2014,Liu2014} describe a problem similar to what we analyze, but they focus on the optimization in a single slot and \emph{not} in the long term. This assumption turns out to be very restrictive in practice. Indeed, in our numerical evaluation we will describe the differences between these two approaches and show that focusing only on a greedy slot-oriented optimization is strongly sub-optimal in the long run. We study the throughput maximization problem and solve it optimally, via the Markov Decision Process (MDP) theory, and approximately, exploiting the results we derived in the optimization section. We explicitly study the trade-offs among battery size, amount of available energy, fading effects and performance. We show how fading and amount of dowlink energy are related and describe how the system changes when the power supply is scarce or abundant.
This work can be considered as a first step to understand the key tradeoffs and optimization problems in a WPCN with finite battery-powered devices.

The paper is organized as follows. Section~\ref{sec:system_model_opt_problem} defines the system model we analyzed and introduces the optimization problem, which is solved in Sections~\ref{sec:opt_solution} and~\ref{sec:app_solution} (optimally and approximately, respectively). We briefly describe the slot-oriented maximization in Section~\ref{sec:traditional}. Section~\ref{sec:numerical_evaluation} presents our numerical results. Finally, Section~\ref{sec:conclusions} concludes the paper.

\section{System Model and Optimization Problem}\label{sec:system_model_opt_problem}

We consider a system composed of three nodes: one Access Point (AP) with Wireless Energy Transfer (WET) capabilities and two devices, namely D$_1$ and D$_2$. Via an RF-WET mechanism, AP recharges the batteries (with finite capacities $B_{1,{\rm max}}$~J and $B_{2,{\rm max}}$~J) of the two devices. It is assumed that AP has an unlimited amount of energy available. The devices use the energy transferred in downlink from the access point to upload data packets.

An approach similar to the ``harvest-then-transmit'' protocol proposed in~\cite{Liu2014} is adopted to keep the devices operational. Under this scheme, time is divided in slots of length $T$ and slot $k$ corresponds to the time interval $[kT,(k+1)T)$. Every slot is divided in two phases:\footnote{Unlike in~\cite{Ju2014}, we choose to consider first the uplink and then the downlink phases in order to more easily track the energy level of the two devices when we set up the MDP formulation in Section~\ref{sec:batteries}.}~
\begin{enumerate}
    \item \textit{uplink} (UP): in the first phase, which lasts for $\tau_1 + \tau_2 \leq T$ seconds, the two devices transmit data to AP in a TDMA fashion using the energy stored in their batteries;
    \item \textit{downlink} (DL): during the second $\tau_{\rm AP} \leq T-\tau_1-\tau_2$ seconds, D$_1$ and D$_2$ harvest the energy transferred from the access point and store it in their batteries.
\end{enumerate}

AP is assumed to have multiple antennas and is able to perform \emph{energy beamforming} in order to split the energy transferred to D$_1$ or D$_2$ during the DL phase, whereas D$_1$ and D$_2$ are assumed to be equipped with an omnidirectional antenna.

\subsection{Uplink Phase}\label{sec:uplink}

At the beginning of a slot, device D$_i$ ($i \in \{1,2\}$) has $B_i \in [0,B_{i,{\rm max}}]$~J of energy stored. In a TDMA fashion, first device $1$ and then device $2$ occupy the channel to transmit data in the uplink channel for $\tau_1$ and $\tau_2$ seconds, respectively. The transmission powers $\rho_1$ and $\rho_2$ and the time allocations $\tau_1$ and $\tau_2$ can change dynamically in every slot and are the control variables of our optimization. We assume that the main source of energy consumption is due to the transmission and therefore neglect the circuitry costs. Note that device D$_i$ is constrained to consume an amount of energy $E_i \triangleq \tau_i \rho_i \leq B_i$ in the upload phase. We also impose the realistic constraint $\rho_i \in [P_{i,{\rm min}},P_{i,{\rm max}}]$ when a transmission is performed. We assume that, in every slot, the devices always have enough data to transmit, \emph{i.e.}, the transmission data queue is always non-empty. This assumption is useful to characterize the  maximum throughput of the system.

According to Shannon's formula, when a power $\rho_i$ is used, the noise power is $\sigma_0^2$ and the uplink channel gain is $h_i$, the transmission rate of device D$_i$ is~
\begin{align}\label{eq:R_rho_h}
    R(\rho_i,h_i) = \log\left(1+\frac{h_i \rho_i}{\sigma_0^2} \right).
\end{align}

Thus, during a single slot, the amount of transmitted data is the time reserved for device D$_i$ multiplied by the transmission rate, $\tau_i R(\rho_i,h_i)$.

The uplink channel is affected by flat fading, which remains constant over the same slot but may change from slot to slot. The channel gain $h_i$ can be expressed as $h_i = \tilde{h}_i \theta_i$, where $\theta_i$ is a random variable which represents the fading and $\tilde{h}_i$ is the average channel gain, obtained by considering the path loss effects as $\tilde{h}_i = h_{0,i} d_i^{-\gamma_i}$. $h_{0,i}$ is the signal power gain at a reference distance of $1$~m, $d_i$ is the distance between D$_i$ and AP expressed in meters, and $\gamma_i$ is the path loss exponent. Note that the device closer to AP experiences, on average, a better channel and spends less energy than the other to transmit the same amount of data (near-far problem).

\subsection{Downlink Phase}

The downlink period lasts for $\tau_{\rm AP} \leq T - \tau_1 - \tau_2$ seconds. During this phase, the access point sends two energy beams to the devices. The sent signal $x_{\rm ET}$ can be seen as the combination of two contributions:~
\begin{align}
    x_{\rm ET} = b_1 s_1 + b_2 s_2,
\end{align}

\noindent where $s_i$ represents the signal carrying energy toward device $i$ and $b_i$ is its directional gain. Since $s_i$ does not carry information, we assume that it is an i.i.d. random variable with zero mean and unit variance. The total power transferred by AP is $\|b_1\|^2 + \|b_2\|^2 \leq Q_{\rm max}$, where $Q_{\rm max} < \infty$ is a technology parameter which represents the maximum power that can be used to transfer energy (\emph{e.g.}, on the order of some watts). The received signal at device D$_i$ is~
\begin{align}
    y_i = \sqrt{g_i} x_{\rm ET} + n_i,
\end{align}

\noindent where $g_i$ is the channel gain from AP to device D$_i$. $n_i$ is the receiver noise, assumed negligible for energy transfer. We assume that  beamforming is perfect, thus $y_i = \sqrt{g_i} b_i s_i$. The corresponding transferred power is~$P_{i,{\rm rc}} = \eta g_i\|b_i\|^2 = \eta g_i Q_i$, where $\eta$ is a constant in $(0,1]$ that models the energy conversion losses at the devices and $Q_i$ is the power sent to device D$_i$, with $Q_1 + Q_2 \leq Q_{\rm max}$. The term $g_i$ can be explicitly written as $g_i = \tilde{g}_i \kappa_i$ in order to consider the flat fading effects, where $\tilde{g}_i = g_{0,i} d_i^{-\delta_i}$ and $\kappa_i$ are defined similarly to $\tilde{h}_i$ and $\theta_i$ in Section~\ref{sec:uplink} ($g_{0,i}$ is the signal power gain at a reference distance of $1$~m and $\delta_i$ is the path loss exponent).
In summary, when a power $Q_i$ is transferred to device $i$, the stored energy is~
\begin{align}\label{eq:C_i}
    C_i = \tau_{\rm AP} \eta Q_i g_{0,i} \left(\frac{1}{d_i}\right)^{\delta_i} \kappa_i.
\end{align}

The channel gain components in uplink $h_1$, $h_2$ and downlink $g_1$, $g_2$ can be assumed equal if the transmission is performed in the same frequency band, which is a common assumption in WPCNs~\cite{Ju2014}. Finally, note that the downlink channel of the user farther from AP is worse (on average), leading to a doubly near-far scenario.

\subsection{Batteries}\label{sec:batteries}

In every slot, the energy level of battery $i$ is updated according to~
\begin{align}\label{eq:battery_evol}
    B_i \leftarrow \min\{B_{i,{\rm max}},B_i - E_i + C_i\}
\end{align}

Note that the arguments of the $\min$ are always non-negative because the energy consumption $E_i$ is chosen such that $E_i \leq B_i$. We also highlight that $C_i$ is a random variable because of the channel fading. The $\min$ operation is used to explicitly consider the effects of finite batteries. The battery evolution depends upon the choices of all parameters $\tau_i$, $\rho_i$, $\tau_{\rm AP}$ and $Q_i$, which are the control variables of our optimization and will be analyzed in the next section.

In order to perform the optimization, we model the system with a discrete Markov Chain (MC). In particular, we discretize the battery of D$_i$ in $b_{i,{\rm max}}+1$ levels, where $b_{i,{\rm max}}$ represents the maximum amount of \emph{energy quanta} that can be stored in the battery and one energy quantum corresponds to $B_{i,{\rm max}}/b_{i,{\rm max}}$~J. There exists a trade-off between the precision of the discrete approximation and the corresponding numerical complexity of the model. In general, if $b_{i,{\rm max}}$ is sufficiently high, the discrete model can be considered as a good approximation of the continuous system. Equation~\eqref{eq:battery_evol} can be rewritten in terms of energy quanta: $b_i \leftarrow \min\{b_{i,{\rm max}},b_i-e_i+c_i\}$. In every slot, only an integer amount of energy quanta $e_i$ can be extracted from the battery. Similarly, only an integer amount of energy quanta can be harvested, thus we define $c_i = \lfloor C_i b_{i,{\rm max}}/B_{i,{\rm max}} \rfloor$ (the \emph{floor} is used to obtain a lower bound to the real performance, whereas an upper bound can be similarly obtained using the ceiling). 
Similarly, if the channel fading is described by a continuous r.v., we discretize it using a finite number of intervals.

In the rest of the paper, the bold notation is used to identify a pair of values, \emph{e.g.}, $\mathbf{a} = (a_1,a_2)$.

\subsection{Optimization Problem}\label{sec:optimization}

We define a \emph{policy} $\mu$ as an action probability measure over the state set, namely $\mathcal{S}$. $\mathcal{S}$ represents all the combinations of battery levels $\mathbf{b}$ and channels $\mathbf{g}$, $\mathbf{h}$. We assume that the policy is computed by a central controller (\emph{e.g.}, the access point), which knows the state of the two batteries $\mathbf{b}$ and the joint channel state $(\mathbf{g},\mathbf{h})$, and distributed among nodes.\footnote{In the cases in which CSI is only partially available, our model is useful to characterize the performance upper bound. A detailed analysis of the partial CSI case is left for future study.} Note that, while estimating the uplink channel is a standard task, downlink channel estimation may be more challenging due to the hardware limitations of the energy receivers. However, by exploiting innovative techniques, \emph{e.g.}, \cite{Xu2014}, it is possible to obtain accurate CSI for the downlink channel as well.

For every state $s = (\mathbf{b},\mathbf{g},\mathbf{h}) \in \mathcal{S}$, $\mu$ defines with which probability an action $a$ is performed. $a$ summarizes the data transmission duration $\boldsymbol{\tau}$, the energy transfer duration $\tau_{\rm AP}$, the transmission powers $\boldsymbol{\rho}$, and the amount of energy $\mathbf{Q}$ to send over the two beams, \emph{i.e.}, $a = (\boldsymbol{\tau},\tau_{\rm AP},\boldsymbol{\rho},\mathbf{Q})$. Formally, $\mu$ defines $\mathbb{P}_\mu(a | s)$, with $\sum_{a \in \mathcal{A}(s)} \mathbb{P}_\mu(a | s) = 1$, where $\mathcal{A}(s)$ is the set of the possible actions in state $s$ (\emph{e.g.}, $\mathcal{A}(s)$ includes the energy constraints imposed by the battery levels). 

For the sake of presentation simplicity, in the next sections we use a deterministic policy $\mu$, \emph{i.e.}, $\mathbb{P}_\mu(a | s)$ is equal to $1$ for $a = \bar{a}_s$ and to $0$ for $a \neq \bar{a}_s$, where $\bar{a}_s$ is an action in $\mathcal{A}(s)$. However, in our numerical evaluation we consider a more general random policy.

Our focus is on the long-term throughput optimization problem. This is suitable for scenarios in which nodes operate in the same position for a sufficient amount of time (\emph{e.g.}, sensors), but can be easily extended to the finite-horizon case with similar techniques. Our goal is to maximize the minimum throughput value reached by both devices in order to increase the QoS. Formally, the reward $G_\mu$ is expressed as~
\begin{align}
    G_\mu &= \min\{G_{1,\mu},G_{2,\mu}\}, \\
    G_{i,\mu} &\triangleq \liminf_{K \to \infty} \frac{1}{K} \sum_{k=0}^{K-1} \mathbb{E}\left[\tau_i R(\rho_i,h_i)\right], \qquad i \in \{1,2\}.
\end{align}

\noindent The expectation is taken with respect to the channel conditions. The maximization process is~
\begin{align} \label{eq:mu_star}
    \mu^\star = \arg \max_\mu \bar{G},
\end{align}

\noindent where $\mu^\star$ is the Optimal Policy (OP). 

\section{Optimal Solution}\label{sec:opt_solution}

In this section, we will show how to solve the problem described in Section~\ref{sec:optimization} and obtain OP. In particular, by exploiting the Markov Decision Process (MDP) theory, the optimization process can be simplified by focusing on the optimization of $\bar{a}_s$ for every fixed $s$ instead of considering the whole function $\mu$, \emph{i.e.}, the optimization can be parallelized (see Bellman's equation in~\cite{Bertsekas2005}). Moreover, we will describe how it is possible to reduce the action $a = (\boldsymbol{\tau},\tau_{\rm AP},\boldsymbol{\rho},\mathbf{Q})$ to a simpler action with only four entries $\tilde{a} = (\tau_{\rm AP},Q_1,\mathbf{e})$.

\subsection{Max-min Problem}\label{sec:max_min}

We now derive a simple technique to deal with the $\max$-$\min$ optimization problem of Equation~\eqref{eq:mu_star}. Indeed, since standard dynamic programming techniques are designed for $\min$ or $\max$ (and not $\max$-$\min$) problems, we recast the problem in a standard form. 

Consider a new optimization problem, similar to the previous one except for the objective function, which becomes $H_\mu(\alpha)$ instead of $G_\mu$:~
\begin{align} \label{eq:H_mu}
    H_\mu(\alpha) = \alpha G_{1,\mu} + (1-\alpha) G_{2,\mu},
\end{align}

\noindent where $\alpha \in [0,1]$ is a constant. Note that the new problem~
\begin{align} \label{eq:mu_star_alpha}
    \mu^\star(\alpha) = \arg \max_\mu H_\mu(\alpha)
\end{align}

\noindent is expressed in a $\max$ form, and thus is easier to solve. If $\alpha = 1$ [$\alpha = 0$], then we are maximizing the performance of device D$_1$ [D$_2$] only and neglecting the other device. 

Name $\mu^\star(\alpha)$ the policy which maximizes $H_\mu(\alpha)$ for a given $\alpha$. Since $\mu^\star(\alpha)$ depends upon $\alpha$, also $G_{1,\mu^\star(\alpha)}$ and $G_{2,\mu^\star(\alpha)}$ implicitly depend upon $\alpha$. It is straightforward to show that $G_{1,\mu^\star(\alpha)}$ [$G_{2,\mu^\star(\alpha)}$] increases [decreases] as $\alpha$ increases. We now want to find the value $\bar{\alpha}$ such that the new problem coincides with the original one. 
Consider the following intuitive result.
\begin{lemma} \label{lemma:G1_G2}
    The optimal solution of Problem~\eqref{eq:mu_star} allocates the same throughput to both users.
\end{lemma}

Therefore, we impose Lemma~\ref{lemma:G1_G2} as design constraint for the new problem and name $\bar{\alpha}$ the value of $\alpha$ at which such condition is satisfied, \emph{i.e.}, $G_{1,\mu^\star(\bar{\alpha})} = G_{2,\mu^\star(\bar{\alpha})}$. Under this condition, we have~
\begin{align}\label{eq:H_G1_G1}
    H_{\mu^\star(\bar{\alpha})}(\bar{\alpha}) = G_{1,\mu^\star(\bar{\alpha})} = G_{2,\mu^\star(\bar{\alpha})}
\end{align}

As a consequence, at $\alpha = \bar{\alpha}$, we obtain $\mu^\star \equiv \mu^\star(\bar{\alpha})$, \emph{i.e.}, OP (solution of~\eqref{eq:mu_star}) coincides with the new policy $\mu^\star(\bar{\alpha})$ which maximizes $H_\mu(\bar{\alpha})$.
This procedure simplifies the numerical optimization because $\mu^\star(\bar{\alpha})$ can be found exploiting standard stochastic optimization algorithms, \emph{e.g.}, the Value Iteration Algorithm (VIA), or the Policy Iteration Algorithm (PIA)~\cite{Bertsekas2005}. 

Practically, the value $\bar{\alpha}$ which satisfies~\eqref{eq:H_G1_G1} can be found with a bisection search as follows. First, arbitrarily fix $\alpha \in [0,1]$ and maximize $H_\mu(\alpha)$ with VIA or PIA. Using the optimal solution, compute $G_{1,\mu^\star(\alpha)}$ and $G_{2,\mu^\star(\alpha)}$. If $G_{1,\mu^\star(\alpha)}$ is greater [less] than $G_{2,\mu^\star(\alpha)}$, then decrease [increase] $\alpha$ and repeat the procedure. The algorithm is repeated until the throughputs of the two nodes are within $\epsilon$ of each other, with $\epsilon$ a sufficiently small constant. In the next, we will (equivalently) deal with $H_\mu(\alpha)$ instead of $G_\mu$.

\subsection{Bellman's Equation Structure}

The most suitable algorithms for solving our problem are VIA or PIA. In the next we describe the \emph{policy improvement step} which is one of the basic operations of both algorithms (see~\cite[Sec.~7.4, Vol.~1]{Bertsekas2005}). 

We define the cost-to-go function associated to state $s$ as $J_s$. The policy improvement step exploits Bellman's equation as follows~
\begin{align}
    J_s &\leftarrow \max_{a \in \mathcal{A}_s}\Big\{ r_\alpha(\boldsymbol{\tau},\boldsymbol{\rho}|\mathbf{h}) +\sum_{s'} \mathbb{P}(s'| s,a) J_{s'} \Big\}, \label{eq:J} \\
    r_\alpha(\boldsymbol{\tau},\boldsymbol{\rho}|\mathbf{h}) &\triangleq {\alpha \tau_{1,\mu} R(\rho_{1,\mu},h_1) + (1-\alpha) \tau_{2,\mu} R(\rho_{2,\mu},h_2)} \nonumber
\end{align}

The probability of going from state $s$ to state $s'$ given the action $a$ can be expressed as~
\begin{subequations}
\begin{align}
    \mathbb{P}(s'| s,a) &\stackrel{(a)}= \mathbb{P}(\mathbf{b}',\mathbf{g}',\mathbf{h}'| \mathbf{b},\mathbf{g},\mathbf{h},a) \label{eq:P_step_1} \\
    &\stackrel{(b)}= \mathbb{P}(\mathbf{b}',\mathbf{g}',\mathbf{h}'| \mathbf{b},\mathbf{g},a) \label{eq:P_step_2} \\
    &\stackrel{(c)}= f(\mathbf{g}',\mathbf{h}')\mathbb{P}(\mathbf{b}'| \mathbf{b},\mathbf{g},a) \label{eq:P_step_3} \\
    &\stackrel{(d)}= f(\mathbf{g}',\mathbf{h}')\mathbb{P}(b_1'| b_1,g_1,a)\mathbb{P}(b_2'| b_2,g_2,a), \label{eq:P_step_4}
\end{align}
\end{subequations}

\noindent where $f(\mathbf{g},\mathbf{h})$ is the pmf of the channel state (note that the randomness is given by the fading components $\theta_i$ and $\kappa_i$ only). $(a)$ holds by definition. $(b)$ holds because the uplink channel does not influence the battery evolution (given the action). $(c)$ holds because the channel is i.i.d. over time and independent of other quantities. The last step holds because the states of the batteries evolve independently in the two devices, given a fixed action. Exploiting Equation~\eqref{eq:C_i} and the MDP formulation, the transition probabilities can be expressed as follows. If $b_i' < b_{i,{\rm max}}$,~
\begin{align}
    \mathbb{P}&(b_i'| b_i,g_i,a) =  \label{eq:P_bi_e}\\
    &\ \chi\Big\{b_i-\tau_i\rho_i +\lfloor \eta g_i \tau_{\rm AP} Q_i b_{i,{\rm max}}/B_{i,{\rm max}} \rfloor = b_i'\Big\}, \nonumber
\end{align}

\noindent otherwise~
\begin{align}
    \mathbb{P}&(b_{i,{\rm max}}'| b_i,g_i,a) =  \label{eq:P_bi_e_max}\\
    &\ \chi\Big\{b_i-\tau_i\rho_i +\lfloor \eta g_i \tau_{\rm AP} Q_i b_{i,{\rm max}}/B_{i,{\rm max}} \rfloor \geq b_{i,{\rm max}}\Big\}. \nonumber
\end{align}

\noindent $\chi\{\cdot\}$ is the indicator function and the \emph{floor} is used to discretize the energy and use the MDP approach. \eqref{eq:P_bi_e}-\eqref{eq:P_bi_e_max} indicate that the battery transitions follow a deterministic scheme (given the action and the state of the system). Intuitively, this happens because the randomness of the channel fading is already included in $g_i$. Therefore, \eqref{eq:J}~ can be reformulated as follows~
\begin{align} \label{eq:J_simplified_det}
    J_s \leftarrow \max_{a \in \mathcal{A}_s}\Bigg\{&\ r_\alpha(\boldsymbol{\tau},\boldsymbol{\rho}|\mathbf{h}) +\sum_{\mathbf{g}',\mathbf{h}'} f(\mathbf{g}',\mathbf{h}')J_{(\mathbf{b}',\mathbf{g}',\mathbf{h}')} \Bigg\},
\end{align}

\noindent with $\mathbf{b}'$ defined according to~\eqref{eq:P_bi_e}-\eqref{eq:P_bi_e_max}. Note that, with this observation, we can avoid to iterate over $\mathbf{b}'$, saving computation time.

Another interesting point is that $\mathbf{b}'$ does not depend upon the particular values of $\boldsymbol{\tau}$ and $\boldsymbol{\rho}$ but only upon their products $\tau_1 \rho_1$ and $\tau_2 \rho_2$. We will use this property in the next section.

\subsection{Variables Reduction}

VIA or PIA requires to focus on the maximization of Equation~\eqref{eq:J_simplified_det} only, which can be formally written as (in this subsection we always refer to a fixed state $s = (\mathbf{b},\mathbf{g},\mathbf{h})$)~
\begin{subequations}
\begin{align}\label{eq:J_simplified_objective}
    \begin{split}
        \max_{\boldsymbol{\tau},\tau_{\rm AP},\boldsymbol{\rho},\mathbf{Q}} \Big\{r_\alpha(\boldsymbol{\tau},\boldsymbol{\rho}|\mathbf{h}) + \Delta(\boldsymbol{\tau}\circ \boldsymbol{\rho},\tau_{\rm AP},\mathbf{Q} | \mathbf{g})\Big\},
    \end{split}
\end{align}
\vspace{-\belowdisplayskip}
\vspace{-\belowdisplayskip}
\vspace{-\abovedisplayskip}
\begin{alignat}{2}
\shortintertext{s.t.:}
    &\tau_i \rho_i \leq B_i = b_i \frac{B_{i,{\rm max}}}{b_{i,{\rm max}}}, \qquad i \in \{1,2\}, \label{eq:tau_rho_B} \\
    &\tau_1 + \tau_2 + \tau_{\rm AP} \leq T, \label{eq:tau1_tau2_tauAP}\\
    &Q_1 + Q_2 \leq Q_{\rm max}, \label{eq:Q_1_Q_2_Q_max}\\
    &\boldsymbol{\tau} \succeq 0,\ \tau_{\rm AP} \geq 0,\ \mathbf{P}_{\rm min} \preceq \boldsymbol{\rho} \preceq \mathbf{P}_{\rm max},\ \mathbf{Q} \succeq 0.\label{eq:constr_great_0}
\end{alignat}
\label{eq:problem_J_simplified}
\end{subequations}

\noindent Constraints~\eqref{eq:tau_rho_B}-\eqref{eq:constr_great_0} represent the set $\mathcal{A}_s$.\footnote{Technically, we should also consider the cases in which $\rho_1 = 0$ and/or $\rho_2 = 0$. However, these are trivial cases that can be easily analyzed separately.} $\succeq$ and $\preceq$ are the component-wise inequalities. $\Delta(\boldsymbol{\tau}\circ \boldsymbol{\rho},\tau_{\rm AP},\mathbf{Q} | \mathbf{g})$ is a quantity that, as the second term in~\eqref{eq:J_simplified_det}, does not depend upon the individual values of $\boldsymbol{\tau}$ and $\boldsymbol{\rho}$ but only on their products ($\circ$ denotes the Hadamard product). This happens because the battery update formulas consider only the overall energy consumption of a device in a slot, that is given by the transmission duration $\tau_i$ multiplied by the transmission power $\rho_i$ (see Equations~\eqref{eq:P_bi_e} and~\eqref{eq:P_bi_e_max}). Without deriving particular properties of $J_s$, the classic procedure to solve~\eqref{eq:problem_J_simplified} is to perform an exhaustive search over all the seven optimization variables. However, the computation may be too demanding\footnote{Note that Problem~\eqref{eq:problem_J_simplified} must be solved for every combination of $\mathbf{b}$, $\mathbf{g}$, $\mathbf{h}$ and for every step of PIA.}  and simpler optimization techniques are required. In particular, in this section we propose a method to simplify the optimization.

First, it can be shown that choosing $Q_1 + Q_2 = Q_{\rm max}$ is optimal (otherwise the available resources would be underused). Similarly, using $\tau_{\rm AP} < T-\tau_1-\tau_2$ is suboptimal. Therefore, without loss of optimality, we can choose $Q_2 = Q_{\rm max}-Q_1$ and $\tau_2 = T - \tau_{\rm AP} - \tau_1$ and avoid to iterate over $Q_2$ and $\tau_2$. Now, fix the products $\boldsymbol{\tau}\circ \boldsymbol{\rho} = \mathbf{E}$, where $E_i$ represents the energy consumed by device D$_i$. In order to solve Problem~\eqref{eq:problem_J_simplified}, we consider the vector $\mathbf{E}$ instead of $\boldsymbol{\tau}$ and~$\boldsymbol{\rho}$.

Given $Q_1$, $\tau_{\rm AP}$, $\mathbf{E}$, the particular values for the duration and the transmission power are extracted by solving the following sub-problem~
\begin{subequations}
\begin{align}
    \begin{split}
        &\max_{\boldsymbol{\tau},\boldsymbol{\rho}}\Big\{ r_\alpha(\boldsymbol{\tau},\boldsymbol{\rho}|\mathbf{h}) + \Delta(\mathbf{E},\tau_{\rm AP},\mathbf{Q} | \mathbf{g}) \Big\},
    \end{split}
\end{align}
\vspace{-\belowdisplayskip}
\vspace{-\belowdisplayskip}
\vspace{-\abovedisplayskip}
\begin{alignat}{2}
\shortintertext{s.t.:}
    &\boldsymbol{\tau}\circ \boldsymbol{\rho} = \mathbf{E}, \label{eq:tau1_rho1}\\
    &\tau_1 + \tau_2 = T - \tau_{\rm AP}, \label{eq:sum_tau_T}\\
    &\boldsymbol{\tau} \succeq 0,\ \mathbf{P}_{\rm min} \preceq \boldsymbol{\rho} \preceq \mathbf{P}_{\rm max},
\end{alignat}
\label{eq:problem_remove_e}
\end{subequations}

\noindent where $\Delta(\mathbf{E},\tau_{\rm AP},\mathbf{Q} | \mathbf{g})$ is a constant term that can be removed from the $\max$ argument. Problem~\eqref{eq:problem_remove_e} can be rewritten as a function of $\tau_1$ only:~
\begin{subequations}
\begin{align}
        \max_{\tau_1}\bigg\{ &\alpha \tau_1 R\left(\frac{E_1}{\tau_1},h_1\right) \\
        &+(1-\alpha)(T-\tau_{\rm AP}-\tau_1) R\left(\frac{E_2}{T-\tau_{\rm AP}-\tau_1},h_2\right)  \bigg\},\nonumber
\end{align}
\vspace{-\belowdisplayskip}
\vspace{-\belowdisplayskip}
\vspace{-\abovedisplayskip}
\begin{alignat}{2}
\shortintertext{s.t.:}
    & \tau_1 \geq \tau_{1,{\rm min}}\triangleq \max &&\left\{ \frac{E_1}{P_{1,{\rm max}}}, T - \tau_{\rm AP} - \frac{E_2}{P_{2,{\rm min}}}\right\}, \label{eq:problem_remove_e_con1}\\
    & \tau_1 \leq \tau_{1,{\rm max}}\triangleq \min &&\left\{ \frac{E_1}{P_{1,{\rm min}}}, T - \tau_{\rm AP} - \frac{E_2}{P_{2,{\rm max}}}\right\}, \label{eq:problem_remove_e_con2}
\end{alignat}
\label{eq:problem_remove_e_tau1}
\end{subequations}

\eqref{eq:problem_remove_e_tau1}~is a uni-dimensional maximization problem which (except in the trivial cases, \emph{e.g.}, $E_1 = 0$ or $E_2 = 0$ or no feasible solutions) can be easily solved by taking the derivative of the reward function, given in the following expression~
\begin{align}
        &\alpha\left(\log \left(1+\frac{h_1}{\sigma_0^2}\frac{E_1}{\tau _1}\right) - \frac{E_1  h_1}{\tau _1 \sigma_0^2 + E_1  h_1}\right)+(1-\alpha) 
         \label{eq:remove_der}\\
        &\times\bigg( \frac{E_2  h_2}{(T-\tau_{\rm AP} -\tau _1)\sigma_0^2 + E_2  h_2} \nonumber\\ 
        &\hspace{3.5cm}-\log \bigg(1+\frac{h_2}{\sigma_0^2}\frac{E_2}{T-\tau_{\rm AP}-\tau _1}\bigg)\bigg), \nonumber
\end{align}

\noindent and setting it to zero. It can be shown that the previous expression has a unique zero in $(0,T-\tau_{\rm AP})$ that corresponds to the optimal value $\tau_{1,{\rm n.c.}}^\star$ of Problem~\eqref{eq:problem_remove_e_tau1} without constraints. The optimal solution of~\eqref{eq:problem_remove_e_tau1}, namely $\tau_1^\star$, can be found as~
\begin{align}
	\tau_1^\star = \max \{ \min \{\tau_{1,{\rm n.c.}}^\star , \tau_{1,{\rm max}}\}, \tau_{1,{\rm min}}\}.
\end{align}

Given $\tau_{\rm AP}$, $\mathbf{E}$ and $\tau_1^\star$, the values of $\tau_2^\star$, $\rho_1^\star$ and $\rho_2^\star$ can be derived from~\eqref{eq:tau1_rho1}-\eqref{eq:sum_tau_T}.

In summary, instead of performing an exhaustive search over seven variables, we just iterate over $\tau_{\rm AP}$, $Q_1$ and $\mathbf{E}$, and recover the other parameters by solving~\eqref{eq:problem_remove_e} and choosing $Q_2 = Q_{\rm max}-Q_1$, $\tau_2 = T-\tau_{\rm AP}-\tau_1$. We also remark that $E_1$, $E_2$ must satisfy $e_i \triangleq E_i b_{i,{\rm max}}/B_{i,{\rm max}} \in \{0,\ldots,b_{i,{\rm max}}\}$.

The previous method is useful to simplify the numerical computation. Moreover, starting from the definitions of $E_1$ and $E_2$, additional insights can be derived and in particular we can exploit the following pruning techniques.

\begin{lemma}
	The energy consumption of \emph{D$_i$} is not decreasing with its uplink channel gain $h_i$.
	
	Formally, consider consider state $(\mathbf{b},\mathbf{g},\mathbf{h}^{\rm (A)})$ and assume that, for fixed $\tau_{\rm AP}$ and $\mathbf{Q}$, the best $\mathbf{e}$ which maximizes~\eqref{eq:J_simplified_objective} is $\mathbf{e}^{\rm (A)}$. Then, for state $(\mathbf{b},\mathbf{g},\mathbf{h}^{\rm (B)})$, with $h_1' \geq h_1$ and $h_2' \leq h_2$, the best $\mathbf{e}$, namely $\mathbf{e}^{\rm (B)}$, satisfies~
	\begin{align}
		e_1^{\rm (B)} \geq e_1^{\rm (A)}, \qquad		e_2^{\rm (B)} \leq e_2^{\rm (A)}
	\end{align}

	\noindent (and similarly by switching the subscripts $1$ and $2$).
\end{lemma}

Intuitively, the previous lemma holds because the reward of Equation~\eqref{eq:R_rho_h} increases with $h_i$. Numerically, this means that, given $\mathbf{e}^{\rm (A)}$, we can avoid to iterate over all the space $(0,\ldots,b_1)\times (0,\ldots,b_2)$ and restrict the optimization space to $(e_1^{\rm (A)},\ldots,b_1)\times (0,\ldots,e_2^{\rm (A)})$ only.

\subsection{Low-SNR Regime}

An interesting and practical case\footnote{Indeed, since the amount of transferred energy is low due to the WET inefficiencies, also the transmission powers are low, leading to a low-SNR scenario.} in which more analytical results can be developed and explained is the low-SNR regime. In this section we provide additional details for such a case. We assume $\frac{\textstyle h_1}{\textstyle \sigma_0^2}\rho_1 \ll 1$ and $\frac{\textstyle h_2}{\textstyle \sigma_0^2}\rho_2 \ll 1$ (low-SNR assumption), therefore $R(\rho_i,h_i) \approx \frac{\textstyle h_i}{\textstyle \sigma_0^2} \rho_i$. In this case, $r_\alpha(\boldsymbol{\tau},\boldsymbol{\rho}|\mathbf{h})$ reduces to $\alpha E_1 \frac{\textstyle h_1}{\textstyle \sigma_0^2} + (1-\alpha) E_2 \frac{\textstyle h_2}{\textstyle \sigma_0^2}$, \emph{i.e.}, it depends only upon the product $\boldsymbol{\tau} \circ \boldsymbol{\rho} = \mathbf{E}$. Therefore, the best choice becomes to use the maximum transmission power $P_{i,{\rm max}}$ and the minimum transmission duration $E_i/P_{i,{\rm max}}$ at both devices. In this way, the system achieves the same reward per slot and maximizes the downlink phase, thus more energy is harvested and stored. As a consequence, once $\mathbf{E}$ is specified, the downlink duration $\tau_{\rm AP}$ is uniquely determined as $\tau_{\rm AP} = T - E_1/P_{1,{\rm max}} - E_2/P_{2,{\rm max}}$.

\subsection{Reducing State Space Complexity}

In a general step of PIA or VIA, given the current policy, the corresponding cost-to-go function $J_s$ has to be computed (policy evaluation step~\cite[Sec.~7.4, Vol.~1]{Bertsekas2005}). This process is challenging when the state space is large.

So far, the state of the system is the tuple $s = (\mathbf{b},\mathbf{g},\mathbf{h})$. However, since $\mathbf{g}$ and $\mathbf{h}$ evolve independently over time, the state space can be reduced to $s = (\mathbf{b})$ only, as follows.
Define a new cost-to-go function~
\begin{align}
    K_{\mathbf{b}} \triangleq \sum_{\mathbf{g},\mathbf{h}} J_{(\mathbf{b},\mathbf{g},\mathbf{h})}.
\end{align}

\noindent $K_{\mathbf{b}}$ substitutes $J_{(\mathbf{b},\mathbf{g},\mathbf{h})}$ in the original problem. Indeed, we can rewrite the policy improvement step as~
\begin{subequations}
\label{eq:K}
\begin{align}
    K_{\mathbf{b}} \leftarrow & \sum_{\mathbf{g},\mathbf{h}} f(\mathbf{g},\mathbf{h}) \max_{a \in \mathcal{A}_{(\mathbf{b},\mathbf{g},\mathbf{h})}} \Big\{ r_\alpha(\boldsymbol{\tau},\boldsymbol{\rho}|\mathbf{h}) + \sum_{s'} \mathbb{P}(s'| s,a) J_{s'}  \Big\} \\
    &= \sum_{\mathbf{g},\mathbf{h}} f(\mathbf{g},\mathbf{h}) \max_{a \in \mathcal{A}_{(\mathbf{b},\mathbf{g},\mathbf{h})}} \left\{ r_\alpha(\boldsymbol{\tau},\boldsymbol{\rho}|\mathbf{h}) + K_{\mathbf{b}'}  \right\},
\end{align}
\end{subequations}

\noindent where $\mathbf{b}'$ is defined according to~\eqref{eq:P_bi_e}-\eqref{eq:P_bi_e_max}.

This procedure further simplifies the numerical computation because 1) it reduces the complexity of the policy evaluation step (there is a lower number of states) and 2) it reduces the number of elementary operations inside the $\max$ operation in the policy improvement step.

\section{Approximate Scheme}\label{sec:app_solution}

Finding the optimal policy is practically feasible only for a relatively small number of discrete values which however corresponds to a rough quantization. Therefore, in this section we propose a method which is based on the characteristics of the original solution but is faster to compute and achieves approximately the same performance of OP. This is particularly useful to characterize the system performance and identify the system trade-offs.

Even with the simplifications introduced in Section~\ref{sec:opt_solution}, the main challenge is to perform the policy improvement step, \emph{i.e.}, solving~\eqref{eq:K} for all system states. 
To manage this problem, several different approximated techniques have been proposed in the literature so far. An interesting idea is to approximate the function $K_{\mathbf{b}}$ with another one simpler to compute. We follow this approach in the remainder of this section, and derive an Approximate Value Iteration Algorithm (App-VIA) (see~\cite[Sec.~6.5]{Bertsekas1996}).

\begin{figure}[t]
  \centering
  \includegraphics[trim = 2mm 0mm 10mm 0mm,  clip, width=.8\columnwidth]{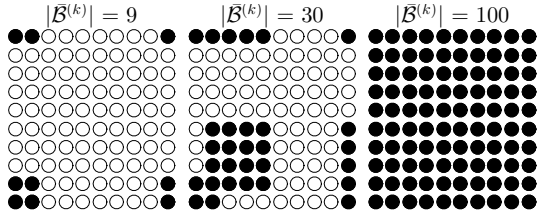}
  \caption{Different sets $\bar{\mathcal{B}}^{(k)}$ when $b_{1,{\rm max}} = b_{2,{\rm max}} = 9$. Rows and columns correspond to $b_1$ and $b_2$, respectively.}
  \label{fig:S_k}
\end{figure}
\subsection{Approximate Value Iteration}

In the classic VIA, the optimal policy is derived by iteratively solving~\eqref{eq:K} until the cost-to-go function converges. In the  approximate approach, we modify every iteration of VIA according to the following two steps:
\begin{enumerate}
    \item compute $K_{\mathbf{b}}^{(k)}$ for every $\mathbf{b} \in \bar{\mathcal{B}}^{(k)}$ performing the policy improvement step (Eq.~\eqref{eq:K}), with $\bar{\mathcal{B}}^{(k)} \subseteq \mathcal{B}$. The superscript $(k)$ denotes the $k$-th iteration of VIA and $\mathcal{B}$ is the set of all battery levels;
    \item interpolate $K_{\mathbf{b}}^{(k)}$ for every $\mathbf{b} \in \mathcal{B}\backslash \bar{\mathcal{B}}^{(k)}$ using the values of $K_{\mathbf{b}}^{(k)}$ computed in the previous step.
\end{enumerate}

The advantage is that the policy improvement is performed only for a subset $\bar{\mathcal{B}}^{(k)}$ rather that for every battery level in~$\mathcal{B}$. See \figurename~\ref{fig:S_k} for a graphical interpretation. A black circle means that $\mathbf{b} \in \bar{\mathcal{B}}^{(k)}$. In the last case, all the battery levels are in $\bar{\mathcal{B}}^{(k)}$, \emph{i.e.}, $\bar{\mathcal{B}}^{(k)} = \mathcal{B}$. In general, $\bar{\mathcal{B}}^{(k)}$ can dynamically change in every step of the algorithm in a deterministic, stochastic or simulation based way. We further discuss our approach in the numerical evaluation section.

We now discuss in more detail the two previous points. The policy improvement step becomes, for every $\mathbf{b} \in \bar{\mathcal{B}}^{(k+1)}$,~
\begin{align}
    &\widehat{K}_\mathbf{b}^{(k+1)} = \sum_{\mathbf{g},\mathbf{h}} f(\mathbf{g},\mathbf{h}) \max_{a \in \mathcal{A}_{(\mathbf{b},\mathbf{g},\mathbf{h})}} \left\{ r_\alpha(\boldsymbol{\tau},\boldsymbol{\rho}|\mathbf{h}) + \widetilde{K}_{\mathbf{b}'}^{(k)}  \right\} \label{eq:K_hat_k_p_1}
\end{align}

\noindent where $\mathbf{b}'$ is defined according to~\eqref{eq:P_bi_e}-\eqref{eq:P_bi_e_max}. $\widehat{K}_\mathbf{b}^{(k+1)}$ represents the approximate value function at step $k+1$ and is defined only in the subset $\bar{\mathcal{B}}^{(k+1)}$, whereas $\widetilde{K}_{\mathbf{b}}^{(k)}$ is such that~
\begin{align} \label{eq:K_tilde_eq_K_hat}
    \widetilde{K}_{\mathbf{b}}^{(k)} = \widehat{K}_\mathbf{b}^{(k)}, \quad & \mbox{if } \mathbf{b} \in \bar{\mathcal{B}}^{(k)}.
\end{align}

\noindent In the second phase of the algorithm, for all $\mathbf{b} \not\in \bar{\mathcal{B}}^{(k)}$, $\widetilde{K}_{\mathbf{b}}^{(k)}$ is derived exploiting~\eqref{eq:K_tilde_eq_K_hat} with an interpolation process or using a mean squares error approximation. In practice, $\widetilde{K}_{\mathbf{b}}(r_k)$ is designed in order to approximate the true function $K_{\mathbf{b}}^{(k)}$. We remark that $\widehat{K}_\mathbf{b}^{(k+1)}$ is defined only in $\bar{\mathcal{B}}^{(k)}$, whereas $\widetilde{K}_{\mathbf{b}}^{(k+1)}$ is defined for every $\mathbf{b} \in \mathcal{B}$.

\subsection{Convergence Properties}
In the following we show that, provided that the approximation $\widetilde{K}_{\mathbf{b}}^{(k)}$ is sufficiently good, the long-term reward of App-VIA is a good approximation of VIA.

First, we introduce the notation $T(\cdot)$ as follows. Define the two sets $\mathcal{K}^{(k)} \triangleq \{K_{\mathbf{b}}^{(k)},\ \forall \mathbf{b} \in \mathcal{B}\}$ and $\widetilde{\mathcal{K}}^{(k)} \triangleq \{\widetilde{K}_{\mathbf{b}}^{(k)},\ \forall \mathbf{b} \in \mathcal{B}\}$. Then, Equations~\eqref{eq:K} and~\eqref{eq:K_hat_k_p_1} can be written as~
\begin{align}
    K_\mathbf{b}^{(k+1)} &= T\left(\mathcal{K}^{(k)},\mathbf{b} \right), \quad \forall \mathbf{b} \in \mathcal{B}, \label{eq:K_T}\\
    \widehat{K}_\mathbf{b}^{(k+1)} &= T\left(\widetilde{\mathcal{K}}^{(k)},\mathbf{b} \right), \quad \forall \mathbf{b} \in \bar{\mathcal{B}}^{(k+1)}, \label{eq:K_tilde_T}
\end{align}

\noindent respectively. Also, assume that the initial configurations are equal, \emph{i.e.}, $\mathcal{K}^{(0)} = \widetilde{\mathcal{K}}^{(0)}$. Note that $K_\mathbf{b}^{(k+1)}$ is evaluated for every $\mathbf{b}$, whereas we compute $\widehat{K}_\mathbf{b}^{(k+1)}$ only in the subset $\bar{\mathcal{B}}^{(k+1)}$.

\begin{propos} \label{propos:Nepsilon}
    After $N$ iterations, the cost-to-go functions of App-VIA and VIA differ by at most $N \epsilon$, \emph{i.e.},\footnote{We adopt the notation $\| \mathcal{K}^{(N)} - \widetilde{\mathcal{K}}^{(N)} \|_\infty \triangleq \max_{\mathbf{b} \in \mathcal{B}} | K_\mathbf{b}^{(N)} - \widetilde{K}_\mathbf{b}^{(N)} |$.}~
    \begin{align} \label{eq:VIA_vs_App_VIA}
        \| \underbrace{\mathcal{K}^{(N)}}_{\rm VIA} - \underbrace{\widetilde{\mathcal{K}}^{(N)}}_{\rm App-VIA} \|_\infty \leq N\epsilon
    \end{align}
    
    \noindent with~
    \begin{align}\label{eq:epsilon_def}
        \epsilon &\triangleq \max_{k=0,\ldots,N-1} \max_{\mathbf{b} \in \mathcal{B}} \left\{\widetilde{K}_\mathbf{b}^{(k+1)} - T \left(\widetilde{\mathcal{K}}^{(k)},\mathbf{b}\right)\right\}
    \end{align}
    \begin{proof}
        See \appendixname~\ref{proof:Nepsilon}.
    \end{proof}
\end{propos}

We first remark that, because of~\eqref{eq:epsilon_def}, Proposition~\ref{propos:Nepsilon} describes a worst case analysis.
$N$ corresponds to the number of iterations of VIA and, in our problem, it can be numerically verified that $N$ is typically small, \emph{e.g.}, $N \approx 10$. The previous proposition provides some bound to the algorithm performance and guarantees convergence, provided that the approximation of $K_\mathbf{b}^{(k+1)}$ is sufficiently good.

\section{Traditional Scheme}\label{sec:traditional}

In the literature, the main focus so far has been on the optimization in a single time slot, which we briefly report in this section for the sake of completeness. In particular, we consider the ``harvest-then-transmit'' scheme, in which all the energy harvested in a slot is used for transmission in the same slot.

If $C_1$ and $C_2$ joules of energy are transferred at the beginning of the slot, in the uplink transmission phase D$_i$ is subject to the following constraint~
\begin{align}
    E_i \leq \min\{C_i,B_{i,{\rm max}}\},
\end{align}

\noindent \emph{i.e.}, it cannot consume more energy than what it received in the same slot nor can it exceed the maximum battery size. The optimization variable is a tuple of $7$ elements. Formally, the optimization problem is~
\begin{subequations}
\begin{align}
    &\max_{\boldsymbol{\tau},\tau_{\rm AP},\boldsymbol{\rho},\mathbf{Q}} \min\{\tau_1 R(\rho_1,h_1),\tau_2 R(\rho_2,h_2)\},
\end{align}
\vspace{-\belowdisplayskip}
\vspace{-\abovedisplayskip}
\begin{alignat}{2}
\shortintertext{s.t.:}
    &\tau_1 + \tau_2 + \tau_{\rm AP} \leq T, \label{eq:trad_t_sum}\\
    &Q_1 + Q_2 \leq Q_{\rm max}, \label{eq:trad_Q_sum}\\
    &\boldsymbol{\tau} \circ \boldsymbol{\rho} \preceq \tau_{\rm AP} \eta \ \mathbf{Q} \circ \boldsymbol{g}, \label{eq:trad_con_sent}\\
    &\boldsymbol{\tau} \circ \boldsymbol{\rho} \preceq \mathbf{B}_{\rm max}, \label{eq:trad_battery_size} \\
    &\boldsymbol{\tau} \succeq 0,\ \tau_{\rm AP} \geq 0,\ \mathbf{P}_{\rm min} \preceq \boldsymbol{\rho} \preceq \mathbf{P}_{\rm max},\ \mathbf{Q} \succeq 0 \label{eq:trad_zero_min_max}
\end{alignat}
\label{eq:trad_problem}
\end{subequations}

As in Section~\ref{sec:opt_solution}, we solve separately the trivial cases ($h_i = 0$, $g_i = 0$, $\rho_i = 0$). The solution of~\eqref{eq:trad_problem} is given in Proposition~\ref{propos:trad_problem}. Constraints~\eqref{eq:trad_battery_size}-\eqref{eq:trad_zero_min_max} identify the \emph{feasible region}. In the following, $-i = 1$ if $i = 2$ and vice-versa.

\begin{propos}\label{propos:trad_problem}
    The optimal $\boldsymbol{\rho}$ (solution of Problem~\eqref{eq:trad_problem}) can be derived as follows (the other parameters are obtained according to Equations~\eqref{eq:trad_tau_AP_sol}-\eqref{eq:trad_Q_i_sol} in Appendix~\ref{proof:trad_problem}).~
    \begin{itemize}
        \item  Name $\rho_i^0$ the solution of~
        \begin{align}
            \eta  g_i Q_{\rm max}+\rho _i = \left(\frac{\sigma_0^2}{h_i} + \rho_i\right) \log \left(1+ \frac{h_i}{\sigma_0^2} \rho_i\right). \label{eq:rho_0}
        \end{align}
        
        \noindent If $\boldsymbol{\rho}^0$ and the corresponding $\boldsymbol{\tau}^0$, $\boldsymbol{Q}^0$, $\tau_{\rm AP}^0$ lie in the feasible region, then $\boldsymbol{\rho}^\star = \boldsymbol{\rho}^0$;
        
        \item otherwise the optimal solution lies on the boundary of the feasible region. 
    \end{itemize}

    \begin{proof}
        See \appendixname~\ref{proof:trad_problem}.
    \end{proof}
\end{propos}

Exploiting the results of the previous proposition, we can derive the optimal reward achieved in a single slot.
By averaging over the channel gains, we obtain the corresponding long-term throughput~
\begin{align} \label{eq:G_sigma}
    G_\sigma \triangleq \sum_{\mathbf{g},\mathbf{h}} f(\mathbf{g},\mathbf{h}) \tau_1^\star R(\rho_1^\star,h_1) =  \sum_{\mathbf{g},\mathbf{h}} f(\mathbf{g},\mathbf{h}) \tau_2^\star R(\rho_2^\star,h_2),
\end{align}

\noindent where $\sigma$ is the \emph{slot-oriented} policy which solves~\eqref{eq:trad_problem}. In the numerical evaluation we will compare $G_\sigma$ and $G_{\mu^\star}$. Note that, differently from $\mu^\star$, the slot-oriented strategy is much simpler to compute but provides lower reward.

\subsection{Low-SNR Regime}
In this section we provide additional details for the low-SNR regime in the case $\boldsymbol{\rho}^\star = \boldsymbol{\rho}^0$. Equation~\eqref{eq:rho_0} can be solved in closed form as~
\begin{align}
    \rho_i^\star =&\ \sqrt{\frac{\eta g_i Q_{\rm max} \sigma_0^2}{h_i}}.
\end{align}

Note that the optimal transmission power of device $i$ depends upon its parameters only. If the downlink channel gain increases, more energy is harvested, therefore a higher transmission power can be used. Interestingly, the better the uplink channel gain $h_i$, the lower the transmission power. The corresponding $Q_i^\star$ can be derived using Equation~\eqref{eq:trad_Q_i_sol}~
\begin{align}
    Q_i^\star =&\ \frac{g_{-i} h_{-i} Q_{\rm max}}{g_1 h_1+g_2 h_2}.
\end{align}

\noindent In order to balance the system performance, $Q_i^\star$ decreases if $g_i$ or $h_i$ increases. In this case, it is better to allocate less resources to the node with a better channel and direct more energy to the other node.

A closed form expression for the reward in a single slot can be obtained. Starting from the equations of $\boldsymbol{\tau}^\star$, $\boldsymbol{\rho}^\star$ and $\mathbf{Q}^\star$, we have~
\begin{align}
    &\tau_1^\star R(\rho_1^\star,h_1) = \tau_2^\star R(\rho_2^\star,h_2) \label{eq:trad_best_tau_R}\\
    &= \frac{\eta  g_1 g_2 \frac{\textstyle h_1 h_2}{\textstyle \sigma_0^2} Q_{\rm max} T}{g_2 h_2\left(\sqrt{\frac{\textstyle \eta g_1 h_1 Q_{\rm max}}{\textstyle \sigma_0^2}}+1\right)+g_1 h_1 \left(\sqrt{\frac{\textstyle \eta g_2 h_2 Q_{\rm max}}{\textstyle \sigma_0^2}}+1\right)}, \nonumber
\end{align}

\noindent which represents the highest reward that can be achieved in a single slot. The long-term reward can be obtained combining the previous expression with~\eqref{eq:G_sigma}, which can be easily solved numerically.

\section{Numerical Results}\label{sec:numerical_evaluation}

We study how the achievable rate changes as a function of the system parameters in different scenarios.
As in~\cite{Ju2014,Liu2014}, we assume channel reciprocity for uplink and downlink, thus $g_i = h_i$ in every slot (however, we remark that our model is general and can be easily adapted to other cases). We consider an exponential random variable with unit mean for $\theta_i$ (Rayleigh fading) to model non line-of-sight links or Nakagami fading with parameter $5$ when a strong line-of-sight component is present. We explicitly consider energy conversion losses by setting $\eta = 0.8$. If not otherwise stated, we use the following parameters $h_{0,1} = h_{0,2} = 1.25 \times 10^{-3}$, $\gamma_1 = \gamma_2 = 2$ (path loss exponents), $\sigma_0^2 = -155$~dBm/Hz (noise power), a bandwidth of $1$~MHz, $T = 500$~ms (slot duration), $Q_{\rm max} = 3$~W (maximum transfer power), $P_{1,{\rm min}} = P_{2,{\rm min}} = 1$~mW and $P_{1,{\rm max}} = P_{2,{\rm max}} = 10$~mW. The battery sizes are important parameters which influence the performance of the system. In particular, since with large batteries the throughput of the system saturates, we choose to focus on the case of small batteries, \emph{i.e.}, $B_{\rm max} \triangleq B_{1,{\rm max}} = B_{2,{\rm max}} \in \{0.1,\ldots,1\}$~mJ~\cite{thinenergy}.

\begin{figure*}[t]
  \includegraphics[trim = 25mm 0mm 15mm 0mm,  clip, width=\textwidth]{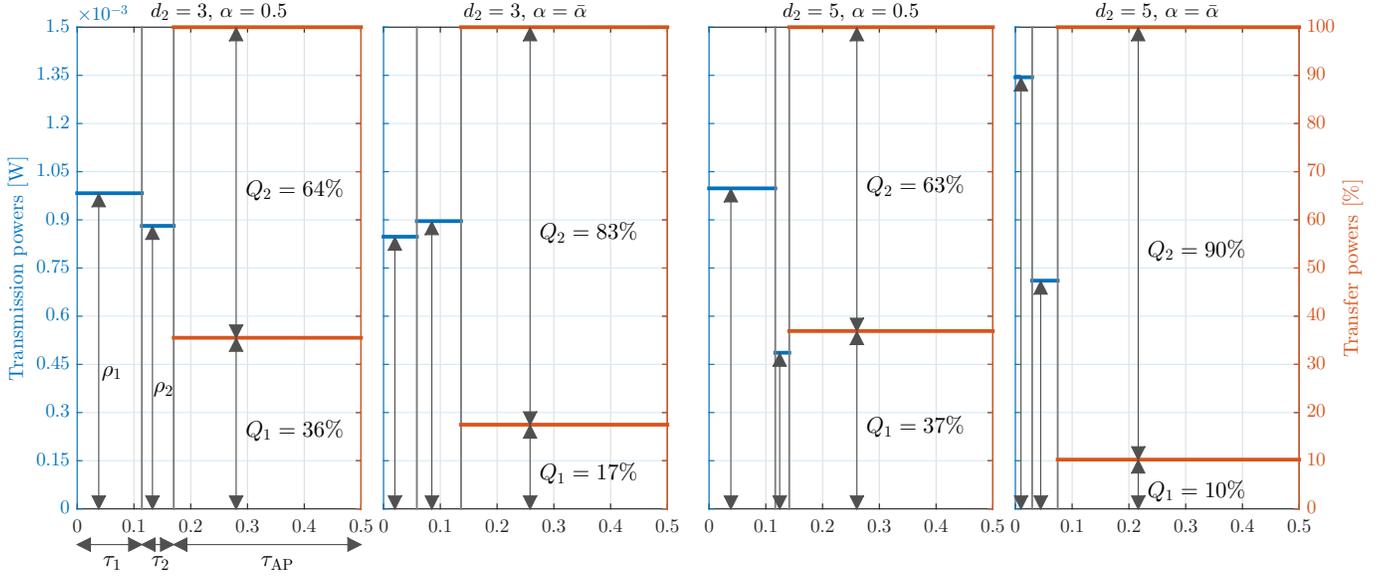}
  \caption{Average transmission powers $\rho_1$, $\rho_2$, transfer powers $Q_1/Q_{\rm max}$, $Q_2/Q_{\rm max}$ and duration $\tau_1$, $\tau_2$, $\tau_{\rm AP}$ with ($\alpha = \bar{\alpha}$) and without ($\alpha = 0.5$) throughput balancing when $d_1 = 1$~m and $d_2 = 3$ or $5$~m.}
  \label{fig:all_iter}
\end{figure*}

In \figurename~\ref{fig:all_iter} we depict the slot division (obtained by averaging all the quantities with the steady-state probabilities) with and without throughput fairness when $d_1 = 1$~m and $d_2 = 3$ or $5$~m. The first figure is obtained by setting $\alpha = 0.5$, \emph{i.e.}, the objective function is the unweighted sum of the rewards of the two devices. Since D$_1$ is closer to AP and experiences, on average, a better channel, it spends more time transmitting. Moreover, even if in $Q_1 < Q_2$, D$_1$ harvests much more energy than D$_2$ on average.
While this scheme achieves the maximum system sum-throughput, it does not ensure fairness. In particular, the throughput of D$_1$ is~$0.88$~Mbps, whereas the throughput of D$_2$ turns out to be only~$0.34$~Mbps. It is also worth noting that D$_2$ does not contribute much to the global performance, but a lot of resources are used to feed it ($Q_2 \gg Q_1$). $Q_1$ is smaller than $Q_2$ because the downlink channel of D$_1$ is better and thus the first device harvests much more energy. When $d_2$ increases as in the third plot of \figurename~\ref{fig:all_iter}, the transmission duration of D$_2$ and its harvested energy become much lower. In this case, D$_2$ is so far from AP with respect to D$_1$ that it is not worth using a lot of resources to increase its throughput.
Instead, the second plot of \figurename~\ref{fig:all_iter} is obtained at the end of the algorithm described in Section~\ref{sec:max_min}, \emph{i.e.}, for $\alpha$ equal to $\bar{\alpha} = 0.91$. With this policy, fairness is achieved and the throughput of the two devices $G_{1,\mu^\star} = G_{2,\mu^\star}$ is~$0.47$~Mbps (which, as expected, results in a smaller sum-throughput than in the unbalanced case). Note that to achieve this situation and to compensate the doubly near-far effect, D$_2$ must receive much more energy and transmit with much more power than D$_1$. This phenomenum is emphasized in the last plot, in which $90\%$ of the transmission power is devoted to D$_2$.

\begin{figure}[t]
  \centering
  \includegraphics[trim = 0mm 0mm 0mm 0mm,  clip, width=1\columnwidth]{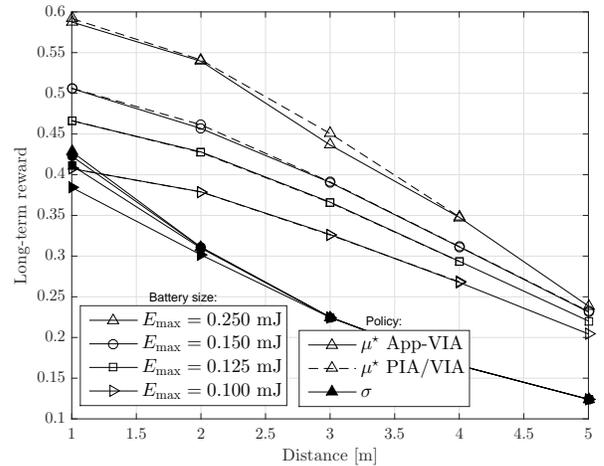}
  \caption{Long-term reward of $\mu^\star$ evaluated with PIA/VIA and App-VIA and of $\sigma$ as a function of $d_1$ when $d_2 = 3$~m and with Rayleigh fading.}
  \label{fig:App_VIA_vs_VIA}
\end{figure}

\begin{figure}[t]
  \centering
  \includegraphics[trim = 0mm 0mm 0mm 0mm,  clip, width=1\columnwidth]{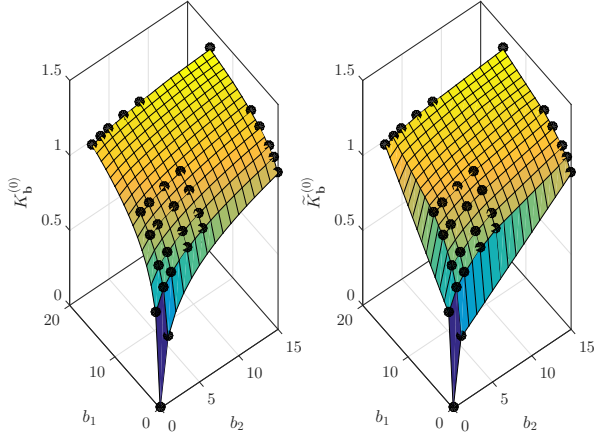}
  \caption{Cost-to-go function $K_{\mathbf{b}}^{(0)}$ (left) and its approximation $\widetilde{K}_{\mathbf{b}}^{(0)}$ (right). Black circles represent $\bar{\mathcal{B}}^{(0)}$.}
  \label{fig:plot_approx}
\end{figure}

We remark that we used a discrete model to approximate the continuous nature of the energy stored in the batteries (see Section~\ref{sec:batteries}), thus $b_{1,{\rm max}}$ and $b_{2,{\rm max}}$ play a key role in the computation of $\mu^\star$. In particular, for larger batteries higher $b_{1,{\rm max}}$ and $b_{2,{\rm max}}$ are required, incurring additional numerical complexity, whereas for small batteries the quantization can be coarser. Nevertheless, even with small batteries, computing the optimal policy $\mu^\star$ with PIA or VIA is a computationally intensive task. Therefore, in the following we present our results using the approximate App-VIA scheme introduced in Section~\ref{sec:app_solution}. To justify the goodness of our approximation, focus on \figurename~\ref{fig:App_VIA_vs_VIA}, where we depict the throughput as a function of the distance $d_1$ for several different battery sizes. It can be seen that App-VIA closely approaches the optimal schemes, especially if the battery sizes are small. In our numerical evaluation we derived $\bar{\mathcal{B}}^{(k)}$ as shown in \figurename~\ref{fig:plot_approx} (see the black circles). The left figure represents the optimal cost-to-go function $K_{\mathbf{b}}^{(0)}$, \emph{i.e.}, Problem~\eqref{eq:K} has been solved for every pair $(b_1,b_2)$, whereas the right plot  represents its approximation $\widetilde{K}_{\mathbf{b}}^{(0)}$ defined in Section~\ref{sec:app_solution}. $\widetilde{K}_{\mathbf{b}}^{(0)}$ is obtained with a linear interpolator.


\begin{figure}[t]
  \centering
  \includegraphics[trim = 0mm 0mm 0mm 0mm,  clip, width=1\columnwidth]{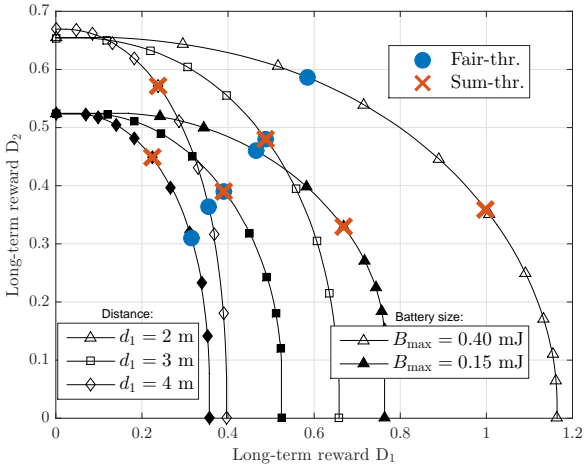}
  \caption{Long-term rewards $G_1$ and $G_2$ of $\mu^\star$ and $\sigma$ when $d_2 = 3$~m and with Rayleigh fading.}
  \label{fig:thr_region}
\end{figure}

\begin{figure}[t]
  \centering
  \includegraphics[trim = 0mm 0mm 0mm 0mm,  clip, width=1\columnwidth]{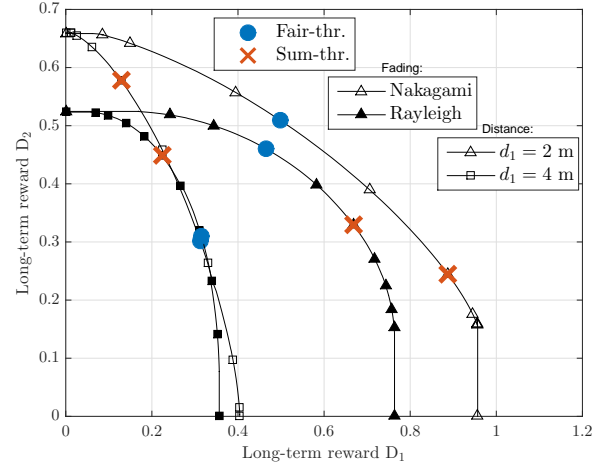}
  \caption{Long-term rewards $G_1$ and $G_2$ of $\mu^\star$ and $\sigma$ with Rayleigh and Nakagami fading when $d_2 = 3$~m and $B_{\rm max} = 0.15$~mJ.}
  \label{fig:thr_region_Nak}
\end{figure}

\figurename~\ref{fig:thr_region} represents the throughput region of D$_1$ and D$_2$, obtained changing $\alpha$ in $(0,1)$. Blue circles represent the fair-throughput optimal points, whereas the red crosses are the sum-throughput optimal points. They coincide only in the symmetric cases $d_1 = d_2$. Otherwise, to balance the system performance, part of the throughput of one of the two devices has be to reduced. Abscissa [ordinate] points are obtained when $\alpha = 1$ [$\alpha = 0$], \emph{i.e.}, D$_2$ [D$_1$] is completely neglected. Similar curves are depicted in \figurename~\ref{fig:thr_region_Nak}, where we compare Rayleigh and Nakagami fading. Even if on average the channel gains are the same in the two scenarios, when a strong line-of-sight component is present (as in Nakagami fading), better performance can be achieved because 1) it becomes easier to predict the future energy arrivals and thus to correctly manage the available energy, and 2) the system approaches the deterministic energy arrivals case, which represents an upper bound for the energy harvesting scenarios~\cite{Biason2015e}.

\begin{figure}[t]
  \centering
  \includegraphics[trim = 0mm 0mm 0mm 0mm,  clip, width=1\columnwidth]{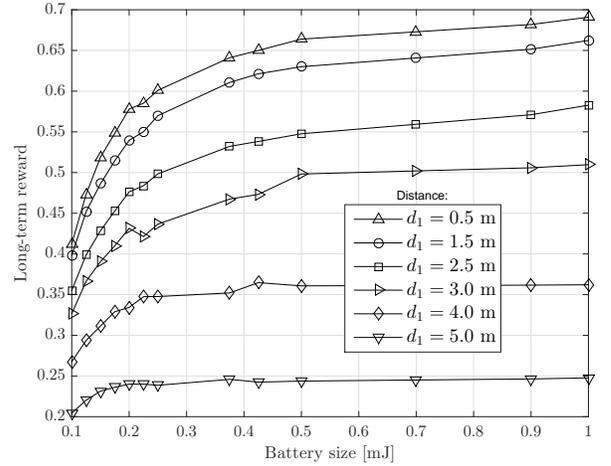}
  \caption{Long-term reward of $\mu^\star$ and $\sigma$ as a function of $E_{\rm max}$ when $d_2 = 3$~m.}
  \label{fig:plot_E_max}
\end{figure}

In \figurename~\ref{fig:plot_E_max} we plot the long-term reward as a function of the battery size of the first device, both for $\mu^\star$. When $B_{\rm max}$ is very small, the batteries represent the system bottleneck because D$_1$ and/or D$_2$ are not able to store and use all the incoming energy. As the battery sizes grow, the performance of the system saturates because the energy available at the access point $Q_{\rm max}$ is limited. The throughput difference between low and high $B_{\rm max}$ is larger when $d_1$ is small because, when the battery of D$_1$ is small, the device is not able to fully exploit its channel potential, which in turn can be fully used with larger batteries. Some artifacts can be noticed (\emph{e.g.}, at $B_{\rm max} = 0.225$~mJ for the curve $d_1 = 3$~m) because we are using App-VIA and not the real optimal policy, whose throughput strictly increases with the battery sizes.

\begin{figure}[t]
  \centering
  \includegraphics[trim = 0mm 0mm 0mm 0mm,  clip, width=1\columnwidth]{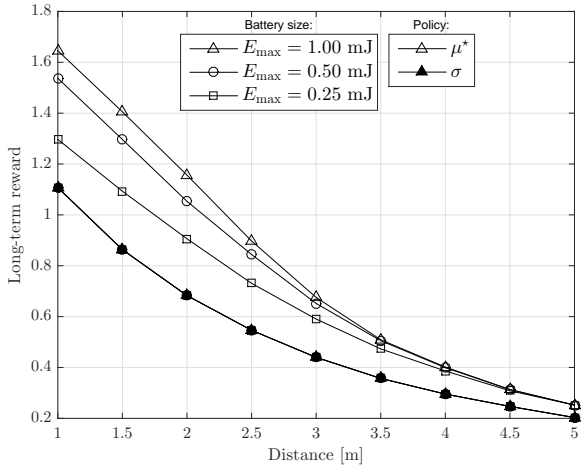}
  \caption{Long-term reward of $\mu^\star$ and $\sigma$ as a function of $d_1$ with high transmission powers when $d_2 = 1$~m.}
  \label{fig:plot_d_high_P_max_d2_1}
\end{figure}

Finally, we describe how the throughput changes as a function of the distance of D$_1$ from AP. \figurename s~\ref{fig:plot_d_high_P_max_d2_1},~\ref{fig:plot_d_high_P_max_d2_3} and~\ref{fig:plot_d_high_P_max_d2_5} are obtained in the ``high transmission power regime,'' \emph{i.e.}, $P_{1,{\rm min}} = P_{2,{\rm min}} = 1$~mW and $P_{1,{\rm max}} = P_{2,{\rm max}} = 10$~mW, whereas \figurename~\ref{fig:plot_d_low_P_max} is determined in the ``low transmission power regime,'' \emph{i.e.}, with $P_{1,{\rm min}} = P_{2,{\rm min}} = 0.01$~mW and $P_{1,{\rm max}} = P_{2,{\rm max}} = 0.5$~mW. When $d_2$ is small (\figurename~\ref{fig:plot_d_high_P_max_d2_1}), the difference between the slot-oriented and the long-term approaches is smaller because a lot of energy is available at the two devices, thus even an inefficient use of it leads to high performance. Instead, as $d_2$ increases (see \figurename~\ref{fig:plot_d_high_P_max_d2_5}), the difference between the two approaches is significant and this supports the need for of a long-term optimization. As expected, in all cases the throughput decreases as $d_1$ increases. This is particularly emphasized when $d_2$ is small because, since it is farther from AP, D$_1$ represents the performance bottleneck. Differently, when $d_2 = 5$~m, D$_2$ is the bottleneck, thus the system performance shows a weak dependence on by the distance of D$_1$ from AP. 
The differences between high and low transmission power regimes can be seen comparing \figurename s~\ref{fig:plot_d_high_P_max_d2_3} and~\ref{fig:plot_d_low_P_max}. It can be seen that with lower transmission powers it is possible to achieve higher rewards. Indeed, in the analyzed scenario the distances are small, thus the uplink SNR is high even for low transmission powers. Therefore, because of the concavity of the reward function in Equation~\eqref{eq:R_rho_h}, with lower transmission powers it may be possible to achieve high throughput while consuming less energy, leading to an overall improvement of the system performance.


\begin{figure}[t]
  \centering
  \includegraphics[trim = 0mm 0mm 0mm 0mm,  clip, width=1\columnwidth]{plot_d_high_P_max.eps}
  \caption{Long-term reward of $\mu^\star$ and  $\sigma$ as a function of $d_1$ with high transmission powers when $d_2 = 3$~m.}
  \label{fig:plot_d_high_P_max_d2_3}
\end{figure}

\begin{figure}[t]
  \centering
  \includegraphics[trim = 0mm 0mm 0mm 0mm,  clip, width=1\columnwidth]{plot_d_high_P_max_d2_5.eps}
  \caption{Long-term reward of $\mu^\star$ and $\sigma$ as a function of $d_1$ with high transmission powers when $d_2 = 5$~m.}
  \label{fig:plot_d_high_P_max_d2_5}
\end{figure}

\begin{figure}[t]
  \centering
  \includegraphics[trim = 0mm 0mm 0mm 0mm,  clip, width=1\columnwidth]{plot_d_low_P_max.eps}
  \caption{Long-term reward of $\mu^\star$ and $\sigma$ as a function of $d_1$ with low transmission powers when $d_2 = 3$~m.}
  \label{fig:plot_d_low_P_max}
\end{figure}


\section{Conclusions}\label{sec:conclusions}

In this work we studied the long-term throughput optimization in a wireless powered communication network composed of an access point and two distributed devices. The system alternates a downlink phase, in which AP recharges the batteries of the nodes via an RF-WET mechanism, and an uplink phase, in which both devices transmit data toward AP in a TDMA fashion. We explained how to solve the long-term throughput maximization problem optimally and approximately while explicitly considering the batteries evolution and the channel state information. We simplified the optimization by exploiting the structure of Bellman's equation. Finally, we compared the long-term approach with the slot-oriented one and noticed that, in order to achieve high performance, the traditional schemes proposed in the literature are strongly sub-optimal.

As part of our future work we would like to 1) extend the model of the system in order to consider partial CSI, storage losses or circuitry costs, 2) compare our results with those obtained using a distributed approach, and 3) extend the long-term optimization to the case with a generic number of nodes.

\appendices

\section{Proof of Proposition~\ref{propos:Nepsilon}}\label{proof:Nepsilon}
The proof is by induction on $N$. If $N = 0$, \eqref{eq:VIA_vs_App_VIA}~holds because $\mathcal{K}^{(0)} = \widetilde{\mathcal{K}}^{(0)}$. Then, assume that~\eqref{eq:VIA_vs_App_VIA} holds for some $N$. The inductive step is as follows~
        \begin{subequations}
        \begin{align}
            &\| \mathcal{K}^{(N+1)} - \widetilde{\mathcal{K}}^{(N+1)} \|_\infty = \max_{\mathbf{b} \in \mathcal{B}} \left| K_\mathbf{b}^{(N+1)} - \widetilde{K}_\mathbf{b}^{(N+1)} \right| \label{eq:proof_1}\\
            & \leq \max_{\mathbf{b} \in \mathcal{B}} \left| K_\mathbf{b}^{(N+1)} - T\left(\widetilde{\mathcal{K}}^{(N)},\mathbf{b} \right) \right| + \label{eq:proof_2}\\
            &\pushright{\max_{\mathbf{b} \in \mathcal{B}} \left| T\left(\widetilde{\mathcal{K}}^{(N)},\mathbf{b} \right) - \widetilde{K}_\mathbf{b}^{(N+1)}  \right| } \nonumber\\
            & \leq \max_{\mathbf{b} \in \mathcal{B}} \left| K_\mathbf{b}^{(N+1)} - T\left(\widetilde{\mathcal{K}}^{(N)},\mathbf{b} \right) \right| + \epsilon \label{eq:proof_3}\\
            & = \max_{\mathbf{b} \in \mathcal{B}} \left| T\left(\mathcal{K}^{(N)},\mathbf{b}\right) - T\left(\widetilde{\mathcal{K}}^{(N)},\mathbf{b} \right) \right| + \epsilon \label{eq:proof_4}\\
            & \leq \max_{\mathbf{b} \in \mathcal{B}} \left| K_\mathbf{b}^{(N)} - \widetilde{K}_\mathbf{b}^{(N)} \right| + \epsilon \label{eq:proof_5}\\
            & = \| \mathcal{K}^{(N)} - \widetilde{\mathcal{K}}^{(N)} \|_\infty + \epsilon \leq N\epsilon + \epsilon = (N+1)\epsilon. \label{eq:proof_6}
        \end{align}
        \end{subequations}
        
        \eqref{eq:proof_1} holds by definition. \eqref{eq:proof_2} exploits the triangular inequality. \eqref{eq:proof_3} uses the hypothesis of the proposition and in particular Definition~\eqref{eq:epsilon_def}. \eqref{eq:proof_4} is by definition of $T$. \eqref{eq:proof_5} is formally proved in the next lemma. \eqref{eq:proof_6} uses the inductive hypothesis. Thus, the proof is concluded by showing the following lemma.
        \begin{lemma}
            Inequality~\eqref{eq:proof_4}-\eqref{eq:proof_5} holds.
            \begin{proof}
            
            	Using Definitions~\eqref{eq:K_T}-\eqref{eq:K_tilde_T}, we obtain~
                \begin{align}
                    &\max_{\mathbf{b} \in \mathcal{B}} \left| T\left(\mathcal{K}^{(N)},\mathbf{b}\right) - T\left(\widetilde{\mathcal{K}}^{(N)},\mathbf{b} \right) \right| \\
                    & = \max_{\mathbf{b} \in \mathcal{B}} \bigg| \sum_{\mathbf{g},\mathbf{h}} f(\mathbf{g},\mathbf{h}) \Big(\max_{a \in \mathcal{A}_{(\mathbf{b},\mathbf{g},\mathbf{h})}} \left\{ r_\alpha(\boldsymbol{\tau},\boldsymbol{\rho}|\mathbf{h}) + K_{\mathbf{b}'}^{(N)}  \right\}  \label{eq:proof_lemma_1} \\
                    & \hspace{3.1cm} - \max_{a \in \mathcal{A}_{(\mathbf{b},\mathbf{g},\mathbf{h})}} \{ r_\alpha(\boldsymbol{\tau},\boldsymbol{\rho}|\mathbf{h}) + \widetilde{K}_{\mathbf{b}'}^{(N)} \} \Big)\bigg|, \nonumber
                \end{align}
                
                \noindent where we recall that $\mathbf{b}'$ is defined according to~\eqref{eq:P_bi_e}-\eqref{eq:P_bi_e_max}. Using the triangular inequality, we work as follows~                
                \begin{align}
                    \eqref{eq:proof_lemma_1} \leq &  \max_{\mathbf{b} \in \mathcal{B}} \bigg| \sum_{\mathbf{g},\mathbf{h}} f(\mathbf{g},\mathbf{h}) \max_{a \in \mathcal{A}_{(\mathbf{b},\mathbf{g},\mathbf{h})}} \Big\{ r_\alpha(\boldsymbol{\tau},\boldsymbol{\rho}|\mathbf{h}) + K_{\mathbf{b}'}^{(N)}  \nonumber  \\
                    &  -  r_\alpha(\boldsymbol{\tau},\boldsymbol{\rho}|\mathbf{h}) - \widetilde{K}_{\mathbf{b}'}^{(N)} \Big\} \bigg|   \\
                    = & \max_{\mathbf{b} \in \mathcal{B}} \bigg| \sum_{\mathbf{g},\mathbf{h}} f(\mathbf{g},\mathbf{h}) \max_{a \in \mathcal{A}_{(\mathbf{b},\mathbf{g},\mathbf{h})}} \Big\{ K_{\mathbf{b}'}^{(N)} - \widetilde{K}_{\mathbf{b}'}^{(N)} \Big\} \bigg|. \label{eq:proof_lemma2}
                \end{align}
                
                Since $\sum_{\mathbf{g},\mathbf{h}} f(\mathbf{g},\mathbf{h}) = 1$, we can substitute the sum with a $\max$ to obtain the upper bound:~
                \begin{align}
                	\eqref{eq:proof_lemma2} \leq \max_{\mathbf{b} \in \mathcal{B}} \ \max_{\mathbf{g},\mathbf{h}} \ \max_{a \in \mathcal{A}_{(\mathbf{b},\mathbf{g},\mathbf{h})}} \Big| K_{\mathbf{b}'}^{(N)} - \widetilde{K}_{\mathbf{b}'}^{(N)} \Big|,
                \end{align}
                
                \noindent which is less than or equal to $\max_{\mathbf{b} \in \mathcal{B}}\left\{K_{\mathbf{b}}^{(N)}-\widetilde{K}_{\mathbf{b}}^{(N)}\right\}$.
            \end{proof}
        \end{lemma}
        
\section{Proof of Proposition~\ref{propos:trad_problem}}\label{proof:trad_problem}
        First, similarly to Lemma~\ref{lemma:G1_G2}, note that the optimal choice leads to $\tau_1^\star R(\rho_1^\star,h_1) = \tau_2^\star R(\rho_2^\star,h_2)$. Also, in order not to underuse the available resources, \eqref{eq:trad_t_sum}-\eqref{eq:trad_Q_sum}~are satisfied with equality. Additionally, we have $\tau_{\rm AP}^\star \eta Q_i^\star g_i \leq B_{i,{\rm max}}$, otherwise the transferred energy would be wasted, which is sub-optimal because the battery could be equivalently filled by reducing $\tau_{\rm AP}$ and increasing $\tau_1$ and/or $\tau_2$ (thus leading to a better solution). Provided this result, it also descends that Constraint~\eqref{eq:trad_con_sent} is always satisfied with equality, \emph{i.e.}, all the energy harvested in a slot is also consumed in the same slot. The problem becomes (we define $R_i \triangleq R(\rho_i,h_i)$ for notation simplicity)~        
\begin{subequations}
\begin{align}
    &\max_{\boldsymbol{\tau},\tau_{\rm AP},\boldsymbol{\rho},\mathbf{Q}} \tau_1 R_1 = \max_{\boldsymbol{\tau},\tau_{\rm AP},\boldsymbol{\rho},\mathbf{Q}} \tau_2 R_2,
\end{align}
\vspace{-\belowdisplayskip}
\vspace{-\belowdisplayskip}
\vspace{-\abovedisplayskip}
\begin{alignat}{2}
\shortintertext{s.t.:}
    &\tau_1 + \tau_2 + \tau_{\rm AP} = T,\\
    &Q_1 + Q_2 = Q_{\rm max},\\
    &\boldsymbol{\tau} \circ \boldsymbol{\rho} = \tau_{\rm AP} \eta \ \mathbf{Q} \circ \boldsymbol{g},\\
    &\tau_1 R_1 = \tau_2 R_2,\\
    &\boldsymbol{\tau} \circ \boldsymbol{\rho} \preceq \mathbf{B}_{\rm max}, \\
    &\boldsymbol{\tau} \succeq 0,\ \tau_{\rm AP} \geq 0,\ \mathbf{P}_{\rm min} \preceq \boldsymbol{\rho} \preceq \mathbf{P}_{\rm max},\ \mathbf{Q} \succeq 0.
\end{alignat}
\end{subequations}

By solving the previous equalities, we can write all the variables as a function of $\boldsymbol{\rho}$~
\begin{subequations}
\begin{align}
    &\max_{\boldsymbol{\rho}} \tau_1 R_1 = \max_{\boldsymbol{\rho}} \tau_2 R_2,
\end{align}
\vspace{-\belowdisplayskip}
\vspace{-\belowdisplayskip}
\vspace{-\abovedisplayskip}
\begin{alignat}{2}
\shortintertext{s.t.:}
    &\boldsymbol{\tau} \circ \boldsymbol{\rho} \preceq \mathbf{B}_{\rm max}, \label{eq:trad_proof_B_max_const}\\
    &\boldsymbol{\tau} \succeq 0,\ \tau_{\rm AP} \geq 0,\ \mathbf{P}_{\rm min} \preceq \boldsymbol{\rho} \preceq \mathbf{P}_{\rm max},\ \mathbf{Q} \succeq 0, \label{eq:trad_proof_P_min_max}
\end{alignat}
\end{subequations}

\noindent where the other parameters are obtained as~
\begin{align}
    &\tau_{\rm AP} = T \frac{g_1 R_1 \rho_2 + g_2 R_2 \rho_1}{\eta g_1g_2Q_{\rm max} (R_1+R_2) + g_1 R_1 \rho_2 + g_2 R_2 \rho_1}, \label{eq:trad_tau_AP_sol}\\
    &\tau_i =  T\frac{\eta g_1 g_2 Q_{\rm max}R_{-i}}{\eta g_1g_2Q_{\rm max} (R_1+R_2) + g_1 R_1 \rho_2 + g_2 R_2 \rho_1}, \label{eq:trad_tau_i_sol}\\
    &Q_i = Q_{\rm max}\frac{R(\rho_{-i},h_{-i})g_{-i}\rho_i}{g_1 R_1 \rho_2 + g_2 R_2 \rho_1}. \label{eq:trad_Q_i_sol}
\end{align}

	To solve the problem, we can take the partial derivatives of $\tau_i R_i$ over $\rho_1$ and $\rho_2$ and setting them to zero (see Expression~\eqref{eq:rho_0}). Using~\cite[Lemma~3.2]{Ju2014} it can be shown that there exists a unique pair of values $(\rho_1^0,\rho_2^0)$ that solves~\eqref{eq:rho_0}, which then corresponds to the global maximum of $\tau_i R_i$. Therefore, if $\boldsymbol{\rho}^0$ and the corresponding $\boldsymbol{\tau}^0$, $\boldsymbol{Q}^0$, $\tau_{\rm AP}^0$ obtained with~\eqref{eq:trad_tau_AP_sol}-\eqref{eq:trad_Q_i_sol} satisfy Constraints~\eqref{eq:trad_proof_B_max_const}-\eqref{eq:trad_proof_P_min_max}, then $\boldsymbol{\rho}^0$ is the optimal solution (unique maximum). Otherwise, the optimal solution must fall on the boundary of the admissible region (since there exists only one stationary point, starting from $\boldsymbol{\rho}^0$, the reward function decreases in every direction).

\bibliography{bibliography}{}
\bibliographystyle{IEEEtran}

\end{document}